\documentclass[12pt,prc,secnumarabic,nofootinbib]{revtex4-1}

\pdfoutput=1

\usepackage{graphicx}

\usepackage{amsmath}
\usepackage{slashed}

\numberwithin{equation}{section}

\DeclareMathOperator{\Tr}{Tr}

\DeclareMathOperator{\sgn}{sign}

\newcommand{\bpsi}{\bar{\psi}}
\newcommand{\hH}{\hat{H}}

\newcommand{\hk}{\hat{k}}
\newcommand{\hq}{\hat{q}}

\newcommand{\vp}{\mathbf{p}}
\newcommand{\vv}{\mathbf{v}}
\newcommand{\vk}{\mathbf{k}}
\newcommand{\vq}{\mathbf{q}}

\newcommand{\pp}{p_{\parallel}}

\newcommand{\rt}{\mathrm{Ret}}
\newcommand{\av}{\mathrm{Adv}}

\newcommand{\ket}[1]{| #1 \rangle}
\newcommand{\bra}[1]{\langle #1 |}

\newcommand{\dg}[1]{#1^{\dagger}}
\newcommand{\gm}[1]{\gamma_{#1}}

\newcommand{\R}{\mathrm{Re}}
\newcommand{\I}{\mathrm{Im}}

\begin{document}

\title{Photon Production from a Quark-Gluon-Plasma at Finite Baryon
Chemical Potential}
\author{Hualong Gervais}
\affiliation{Department of Physics, McGill University, Montreal, Canada}
\author{Sangyong Jeon}
\affiliation{Department of Physics, McGill University, Montreal, Canada}


\begin{abstract}
We compute the photon production of a QCD plasma at leading order in the
strong coupling with a finite baryon chemical potential. Our approach
starts from the real time formalism of finite temperature field theory. We
identify the class of diagrams contributing at leading order when a finite
chemical potential is added and resum them to perform a full treatment of
the Landau-Pomeranchuk-Migdal (LPM) effect similar to the one performed by Arnold, Moore, and Yaffe at
zero chemical potential. Our results show that the contribution of
$2\mapsto 3$ and $3\mapsto 2$ processes grows
as the chemical potential grows.
\end{abstract}


\maketitle

\section{Introduction}\label{sect:intro}

Heavy ion collision experiments at RHIC typically create a medium where
the
net baryon density is
non-vanishing~\cite{Andronic:2005yp,BraunMunzinger:2011ze}.
As we enter a new era of precision
measurements, it is therefore important to consider the effect of
a
non-vanishing baryon chemical potential on the thermal photon yield of the
quark gluon plasma.
Previously, the effect of non-zero chemical potential
in $2\leftrightarrow 2$ processes
was studied in
\cite{Dumitru:1993us,Traxler:1994hy,Dutta:1999dy,He:2005yb}.
The complete leading order
calculation of photon production including the effect of collinear
enhancement in the $2\to 3$ and $3\to 2$ cases
was first carried out by Arnold, Moore, and
Yaffe (AMY) starting from first principles within the framework of quantum
field theory at finite temperature~\cite{Arnold:2001ba}.
The analysis of \cite{Arnold:2001ba}, however, is mostly carried out at
zero chemical potential.
The effect of adding a chemical potential was mentioned, but a detailed
analysis was not fully performed.

The chemical potential will modify the quark and gluon
self-energies as well as change the statistical factors,
thereby potentially modifying the power counting analysis of AMY.
In this paper, we will precisely determine under which circumstances the
power
counting must be modified to account for the presence of a chemical
potential. We also explore numerically the consequences of including a
chemical potential
in the photon production.

As in the previous cases, the most convenient basis to work in when
analyzing the
parametric sizes of diagrams is the Keldysh or $r,a$ basis whereas
the rates are most conveniently written in the usual $1,2$ basis.
The addition of a finite chemical potential also changes the way in which
we
switch from the $1,2$ basis to the $r,a$ basis, and we will carefully
explain
the required changes.

The organization of this paper is as follows. In section \ref{sect:G1122},
we briefly recall the structure of perturbation theory in the real time
formalism and outline the basic formula giving photon production in terms
of Feynman diagrams. In section \ref{sect:heinz_and_wang}, we show how to
generalize a result of Heinz and Wang~\cite{Wang:1998wg} that allows one
to
express Green functions in the $1,2$ basis in terms of a reduced set of
Green functions in the $r,a$ basis. In section
\ref{sect:hard_thermal_loops}, we show how a finite chemical potential
alters quark and gluon thermal masses. Then, in section
\ref{sect:power_counting}, we perform a power counting analysis to
determine which diagrams contribute at leading order in the strong coupling $g$ and resum these in
section \ref{sect:resummation}. Finally, some numerical results are shown
in section \ref{sect:numerical_results}.


\section{Diagrammatic Approach to Calculating Photon
Production}

\subsection{Perturbation Theory at Finite Temperature and Chemical Potential}\label{sect:G1122}

We start by briefly outlining the structure of real time perturbation
theory at finite chemical potential. Since we are interested in a QCD
plasma, our starting point is the QCD Lagrangian:
\begin{align}
\mathcal{L} &=
\sum_{f}\bpsi_{f}(\slashed{\partial}-M-g\slashed{A}_aT^{a})\psi_{f}
-\frac{1}{4}F_{\mu\nu}^{a}F_{a}^{\mu\nu}
\end{align}
where the sum is over the $N_f$ fermion flavors and the gauge group is
$\mathrm{SU}(N_c)$ with $N_c = 3$.\footnote{Our metric convention is
$g_{\mu\nu} = {\rm diag}(1,-1,-1,-1)$.}
This Lagrangian has a conserved charge
$\hat{Q}\equiv \sum_f\int d^3x\psi^\dagger_f(x)\psi_f(x)$
which is equal to the net
fermion (quark) number. The density operator describing the
grand-canonical
ensemble is therefore $e^{-\beta(\hH-\mu\hat{Q})}$.
In the imaginary time formalism, one
can show that this changes the Matsubara frequencies from $i\omega_n$ to
$i\omega_n+\mu$~\cite{Kapusta:2006pm,LeBellac00}.
The chemical potential here is hence the {\em quark} chemical potential
which is 1/3 of the baryon chemical potential.

To connect the imaginary and real time formalisms, one defines the
retarded
and advanced propagators:
\begin{align}
G_{\rt}(x)&=i\theta(t)\langle \{\psi(x),\bpsi(0)\}\rangle_{\beta,\mu}\\
G_{\av}(x)&=-i\theta(-t)\langle \{\psi(x),\bpsi(0)\}\rangle_{\beta,\mu}
\end{align}
The anticommutators above are replaced with commutators for gauge fields.
By using the spectral representation of imaginary time propagators
(for instance, see \cite{Kapusta:2006pm,LeBellac00})
one can show that the retarded propagator is
obtained by analytically continuing the Matsubara propagator through the
prescription $i\omega_n+\mu\mapsto p^0+i\epsilon$ and the advanced
propagator
from the continuation $i\omega_n+\mu\mapsto p^0-i\epsilon$, where $p^0$ is
an
arbitrary real number.

In the real time formalism, one is forced to double the degrees of freedom
because when constructing a generating functional from
the partition function, the time integration contour has to traverse the
real time axis once forward and backwards
(Schwinger-Keldysh Closed Time Path) \cite{Keldysh:1964ud}. The first set of fields
$\psi_{1}$, $\bpsi_{1}$, and $A^{a,\mu}_{1}$ corresponds to fields with a
time argument on the forward directed part of the contour and conversely
the set of fields $\psi_{2}$, $\bpsi_{2}$, and $A^{a,\mu}_{2}$ corresponds
to fields with time arguments on the backwards directed part.

For computational purposes, it is sometimes more convenient to use another
basis
than the $1,2$ basis above. One such basis is the $r,a$ or
Keldysh basis \cite{Keldysh:1964ud,Chou:1984es},
defined as follows:
\begin{align}
\label{eq:def_ra_basis}
\varphi_r=\frac{\varphi_1+\varphi_2}{2}\\
\varphi_a=\varphi_1-\varphi_2
\end{align}
where $\varphi$ denotes any of our fields. By using the above algebraic
relation and the spectral representation of propagators in the $r,a$
basis,
one can show that for fermions:
\begin{align}
G_{rr}(P)&=\left(\frac{1}{2}-n_{f}(p^0-\mu)\right)\rho(P)\\
G_{ar}(P)&=G_{\av}(P)\\
G_{ra}(P)&=G_{\rt}(P)\\
G_{aa}(P)&=0 \label{eq:Gaa}
\end{align}
The above allows us to compute real time propagators by analytically
continuing imaginary time ones.
{}From now on, we use capital letters for 4-momenta and lower case letters
to
denote the magnitude of the 3-momenta.
For instance, $P = (p^0, {\bf p})$ in the above expressions and $p = |{\bf
p}|$.

To tie this formalism to the problem of photon production, we recall the
standard formula that relates the emissivity of the plasma to a Wightman
current-current correlator \cite{Kapusta:2006pm,LeBellac00}
\begin{align}
\frac{d\Gamma_\gamma}{d V}=\frac{d^3 \vk}{(2\pi)^3 2|\vk|} \sum_{a=1,2}
\epsilon_\mu^{(a)}(K) \epsilon_\nu^{(a)}(K) W^{\mu\nu}(K)
\end{align}
where we have defined the Wightman current-current correlator by:
\begin{align}
W^{\mu\nu}(K) = \int d^4 x\; e^{iK\cdot x} \langle j^\mu(0) j^\nu(x)
\rangle_{\beta,\mu}\label{eq:current_current}
\end{align}
Wightman functions are most naturally expressed in the $1,2$ formalism of
the
real time theory:
\begin{align}
d\Gamma_\gamma &=
\frac{\alpha_{EM}}{\pi^2}\frac{d^3\vk}{|\vk|}\int\frac{d^4P_1}{(2\pi)^4}\int\frac{d^4 P_2}{(2\pi)^4}\frac{((p_{1\parallel}+k)^2+p_{1\parallel}^2)}{2p_{1\parallel}(p_{1\parallel}+k)}(P_{1\perp}\cdot P_{2\perp})\times\nonumber\\
&\quad\times G_{1122}(-P_1,K+P_1,-K-P_2,P_2)
\end{align}
where $G_{1122}$ is a fermionic four-point function where two external vertices are
of the `1' types
and the others are of the `2' types. The ``parallel'' component of $\vp$ is
always defined relative to the fixed direction of the emitted photon
momentum $\vk$.

\begin{figure}[t]
\centering
\includegraphics[width=0.7\textwidth]{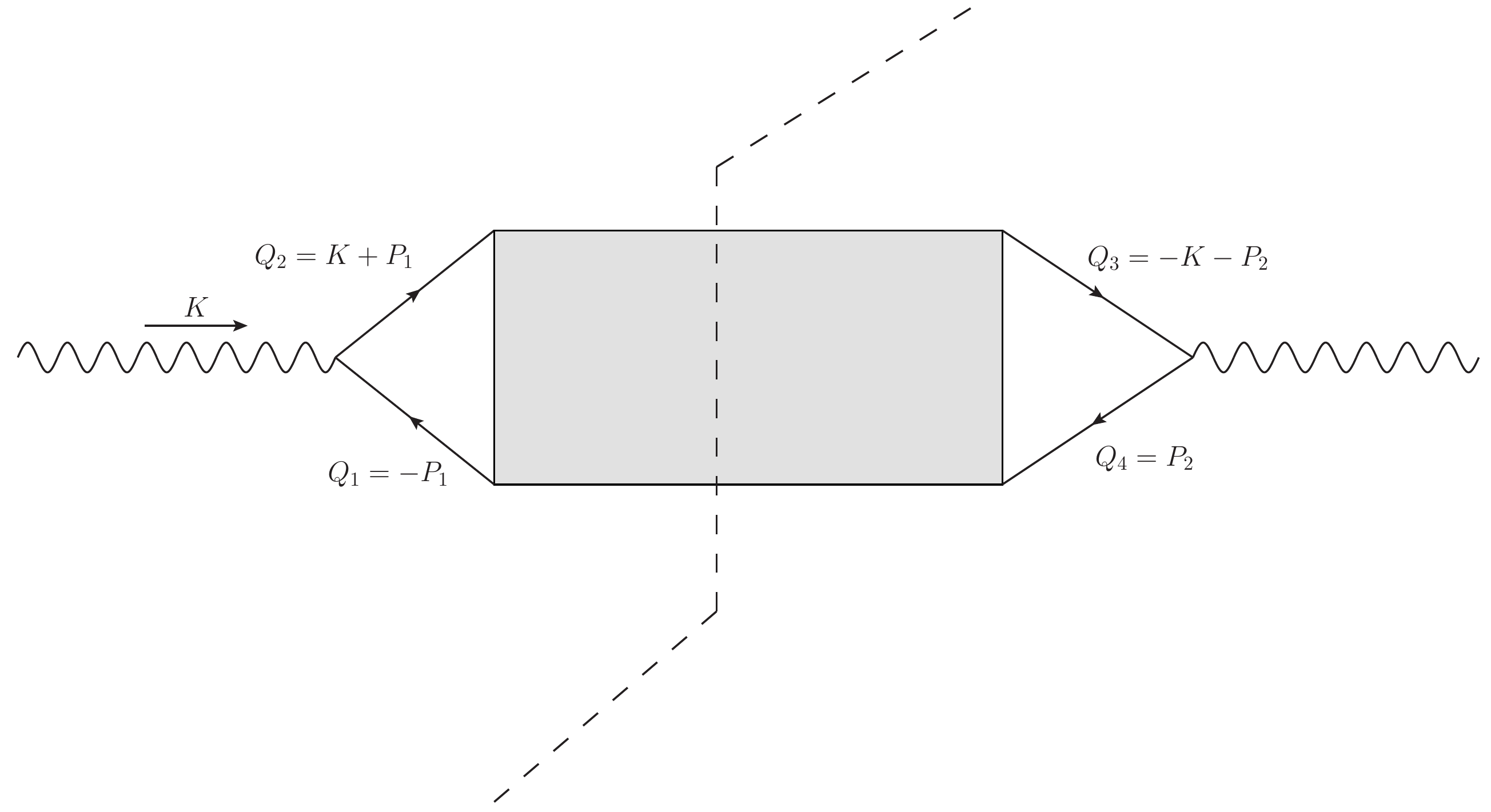}
\caption{The momentum assignments for computing the Wightman function.
Arrows on fermion lines going into the shaded box correspond to insertions
of the particle operator and those
coming out of it corrrespond to insertions of the antiparticle operator.
All momenta flow into the shaded box.}
\label{fig:momenta_assignments}
\end{figure}

A schematic diagram for $d\Gamma_\gamma$ is shown in
figure~\ref{fig:momenta_assignments}.
Our convention is that fermion momenta flow into the shaded box.
Arrows going into the shaded box correspond to
insertions of the particle operator $\psi$, and arrows coming out of it
correspond to antiparticle insertions $\bpsi$.


\subsection{Going from the $1,2$ to the $r,a$ Basis}
\label{sect:heinz_and_wang}

For power counting and actual computations,
the $r,a$ basis is more convenient than the $1,2$ basis.
By using the algebraic relation
between fields in the $1,2$ and $r,a$ bases
(c.f.~Eq.(\ref{eq:def_ra_basis})),
we can express $G_{1122}$ as a
linear combination of the sixteen possible four-point functions in the
$r,a$ basis. The result is:
\begin{align}
G_{1122}&=G_{rrrr}+\frac{1}{2}G_{rarr}+\frac{1}{2}G_{arrr}+\frac{1}{4}G_{aarr}\nonumber\\
&-\frac{1}{2}G_{rrar}-\frac{1}{2}G_{rrra}+\frac{1}{4}G_{rraa}-\frac{1}{4}G_{rara}\nonumber\\
&-\frac{1}{4}G_{arra}-\frac{1}{4}G_{raar}-\frac{1}{4}G_{arar}+\frac{1}{8}G_{araa}\nonumber\\
&-\frac{1}{8}G_{aara}+\frac{1}{8}G_{raaa}-\frac{1}{8}G_{aaar}+\frac{1}{16}G_{aaaa}
\end{align}
One can show that in general, $G_{aa...a}(1,2,...,n)=0$, and hence we can
immediately eliminate one of the sixteen terms in the above -- See for
example \cite{Chou:1984es}.
In Ref.\cite{Wang:1998wg}, Wang and Heinz showed that
the remaining 15 terms can be re-expressed in terms of
seven four-point functions  and their complex conjugates.

The authors of Ref.\cite{Wang:1998wg}
only considered the case of a real scalar field.
However, their derivation works equally well for fermions -- See appendix
A. Hence
the following results from the reference still holds
\begin{align}
G_{1122}&= \alpha_1 G_{aarr}+\alpha_2 G_{aaar}+\alpha_3 G_{aara}+\alpha_4
G_{araa}+\alpha_5 G_{raaa}+\alpha_6 G_{arra}+\alpha_7 G_{arar}\nonumber\\
&+\beta_1\bar{G}^{*}_{aarr}+\beta_2\bar{G}^{*}_{aaar}
+\beta_3\bar{G}^{*}_{aara}+\beta_4\bar{G}^{*}_{araa}
+\beta_5\bar{G}^{*}_{raaa}+\beta_6\bar{G}^{*}_{arra}+\beta_7\bar{G}^{*}_{arar}\label{eq:change_basis}
\end{align}
with the coefficients $\alpha_i$ and $\beta_i$ composed of Fermi-Dirac
distribution functions instead of Bose-Einstein distribution functions and
$\bar{G}$ denoting the charge conjugate ($\bpsi\leftrightarrow \psi$) of
$G$.
Here all $G_{\cdots}(Q_1,Q_2,Q_3,Q_4)$ are functions of 4 momenta.
Our convention is that the momenta $Q_2$ and $Q_4$ correspond to the
insertion of $\psi$ and $Q_1$ and $Q_3$ correspond to
the insertion of $\bar\psi$.

This expression greatly reduces the number of diagrams we
need to estimate. As a matter of fact, what emerges from our power
counting
analysis is that only $G_{aarr}$ contributes to photon production at
leading order. This is because only this labeling gives rise to pinching
poles as we will
explain shortly.
Thus, we only need to know the coefficients $\alpha_1$ and
$\beta_1$.  These are given by
\begin{align}
\alpha_1 &= n_f(q_1^0+\mu)n_f(q_2^0-\mu)\label{eq:wang_heinx_gen_1}\\
\beta_1 &=
-(1-n_f(q_3^0+\mu))(1-n_f(q_4^0-\mu))\frac{1-n_f(q_1^0+\mu)-n_f(q_2^0-\mu)}{1-n_f(q_3^0+\mu)-n_f(q_4^0-\mu)}\label{eq:wang_heinx_gen_2}
\end{align}
This is proven in Appendix A.
One can intuitively understand the signs of $\mu$ in the above by
recalling
that $n_f(E-\mu)$ is the distribution function for particles while
$n_f(E+\mu)$ is the distribution function for antiparticles.
Since $Q_1$
and $Q_3$ correspond to antiparticle insertions, they must be associated
with the distribution $n_f(q_i^0+\mu)$, and conversely for $Q_2$ and
$Q_4$.

As we will discuss in section \ref{sect:power_counting}, all gluon exchange
momenta must be soft at leading order in the strong coupling $g$. Therefore, we have that $P_1\simeq
P_2$ in figure \ref{fig:momenta_assignments}. Hence, $Q_1\simeq -Q_4$ and
$Q_2\simeq -Q_3$
and consequently, $\beta_1\simeq n_f(q_1^0+\mu)n_f(q_2^0-\mu)=\alpha_1$.
Further, one can verify that
$\bar{G}_{aarr}(-P_1,K+P_1,-K-P_2,P_2)=G_{aarr}(-P_1,K+P_1,-K-P_2,P_2)$
--See appendix B.
Therefore, only the real part of $G_{aarr}$ is
relevant to photon production at leading order.


\subsection{Self-Energies at Finite Chemical
Potentials}\label{sect:hard_thermal_loops}

When we compute the Wightman correlator to derive the photon production,
we
need to do a power counting analysis to identify all leading order
diagrams. In this analysis, the appearance of ``pinching poles'' makes it
crucial to resum the thermal self-energies into the quark and gluon
propagators. Therefore, we need to know how the presence of a chemical potential affects
the self-energies of quarks and gluons. This is well-known and dates back to the original paper
by Braaten and Pisarski on hard thermal loops \cite{Braaten:1989mz}.

In order to keep our work self-contained, we review the derivation of self-energies with a full inclusion of a chemical potential
in appendix C. In this section, we simply quote the final results.

The gluon polarization tensor at finite chemical potential $\mu$ takes the form:
\begin{align}
\Pi_{\mu\nu}(P) &=
m_D^2\left(-\delta_{0}^{\mu}\delta_{0}^{\nu}+\int\frac{d\Omega}{4\pi}\frac{p^0v_\mu v_\nu}{p^0-\vv\cdot\vp+i\epsilon}\right)\label{eq:HTL_gluons}
\end{align}
where in the above, $m_D^2=g^2\left(\frac{N_f T^2}{6}
+\frac{N_c T^2}{3}+\frac{N_f\mu^2}{2\pi^2}\right)$ is the Debye mass, and
$v_\mu=(1,\vv)$ with $\vv=\vk/|\vk|$.

Similarly, the quark self-energy at finite chemical potential is:
\begin{align}
\Sigma_{T}(P)
&=\frac{g^2
C_{2}(F)}{8}\left(T^2+\frac{\mu^2}{\pi^2}\right)\int\frac{d\Omega}{4\pi}\frac{\slashed{v}}{p^0-\vv\cdot\vp+i\epsilon}
\end{align}

The above two results show that the only effect of the chemical potential is to shift the dependence of
self-energies on $T^2$ by a $\mu^2$ term. As we will see, the consequence of this is that in carrying
out our power counting analysis, we need not alter the arguments of Arnold, Moore, and Yaffe as long as
$\mu \leq O(T)$.


\section{Power Counting with a Finite Chemical
Potential}\label{sect:power_counting}

When evaluating the Wightman function, there are two regions of the
spatial
part of the loop momentum integration that are of interest.
\begin{itemize}
\item The non-collinear region: ${\bf p}_{1,\perp}$
and ${\bf p}_{2,\perp}$ are both $O(T)$.
\item The near-collinear region: ${\bf p}_{1,\perp}$
or ${\bf p}_{2,\perp}$ is $O(gT)$ or less.
\end{itemize}

The non-collinear region corresponds to the contributions of the basic
$2\mapsto 2$ processes treated by Kapusta et al. \cite{Kapusta:1991qp} and
Baier et al. \cite{Baier:1991em}. The near-collinear region will have both
pinching pole and near-collinear enhancements. It corresponds to the
contribution of the bremsstrahlung and inelastic pair annihilation
processes modified by the Landau-Pomeranchuk-Migdal (LPM) effect.

\begin{figure}[t]
\centering
\includegraphics[width=0.4\textwidth]{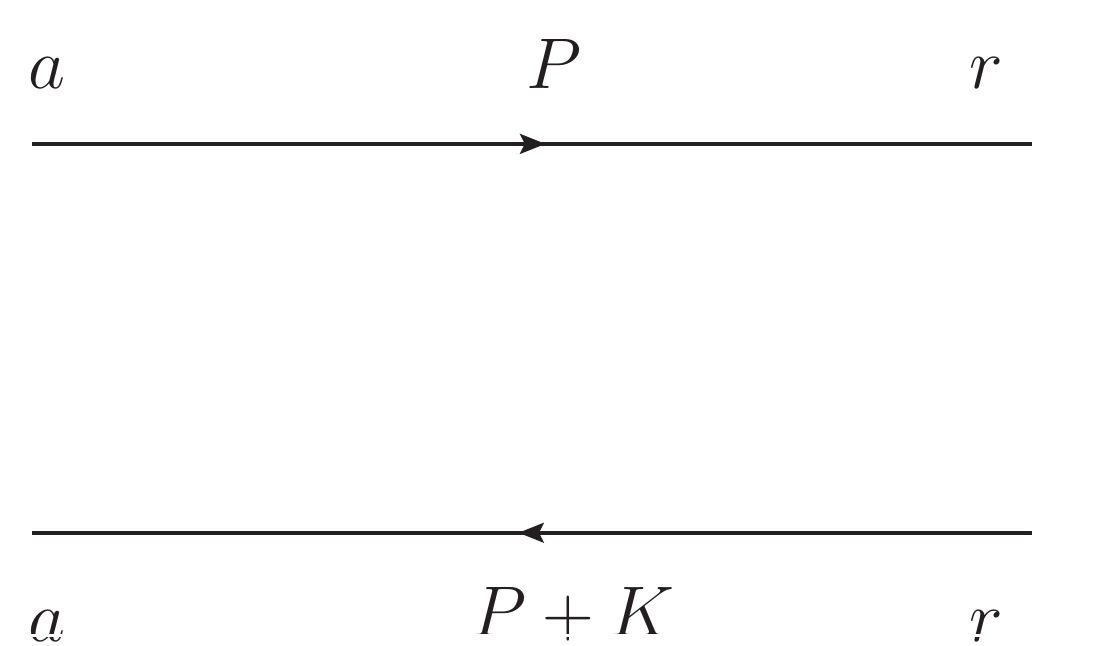}
\caption{The pair of fermionic propagators carry momenta which differ by
$K$. We assume that $K$ is nearly collinear with $P$ and that the other
parts of the diagram are essentially constant over the region of width
$g^2T$ centered around the pinching poles of the integrand.}
\label{fig:pinching_poles}
\end{figure}

We briefly review how pinching pole enhancements arise.
As will be shortly shown,
the leading order
diagrams for photon radiation all contain a pair of fermionic propagators
as shown in
figure \ref{fig:pinching_poles}.
The spatial part $\vp$ of the loop
momentum $P$ is assumed to be nearly collinear with $\vk$. In other words,
$\vp_{\perp}\sim gT$.
Given that we impose the collinearity of $\vp$ and $\vk$, we may focus just on
the frequency integral
\begin{align}
&\int\frac{dp^0}{2\pi}G_{\mathrm{Adv}}(K+P)G_{\mathrm{Ret}}(P)
=\nonumber\\
&\quad\int\frac{dp^0}{2\pi}\frac{1}{\left[\left(p^0+\frac{i\Gamma_\vp}{2}\right)^2
-E_\vp^2\right]\left[\left(p^0+k^0-\frac{i\Gamma_{\vp+\vk}}{2}\right)^2-E_{\vp+\vk}^2\right]}
\label{eq:freq_int}
\end{align}
where we have absorbed the Dirac matrix structures into the vertices.
In Eq.(\ref{eq:freq_int}),
$\Gamma_{\vp}=\I(\Sigma(P))/(2E_\vp)$ is the decay width of
the quark generated by the imaginary part of the self energy. To leading
order in $g$, it can be replaced by its asymptotic value
$\Gamma=\mathrm{lim}_{\vp\rightarrow\infty}(\Gamma_\vp)$
(\cite{Flechsig:1995ju,Kraemmer:1989dr}).

\begin{figure}[t]
\centering
\includegraphics[width=0.7\textwidth]{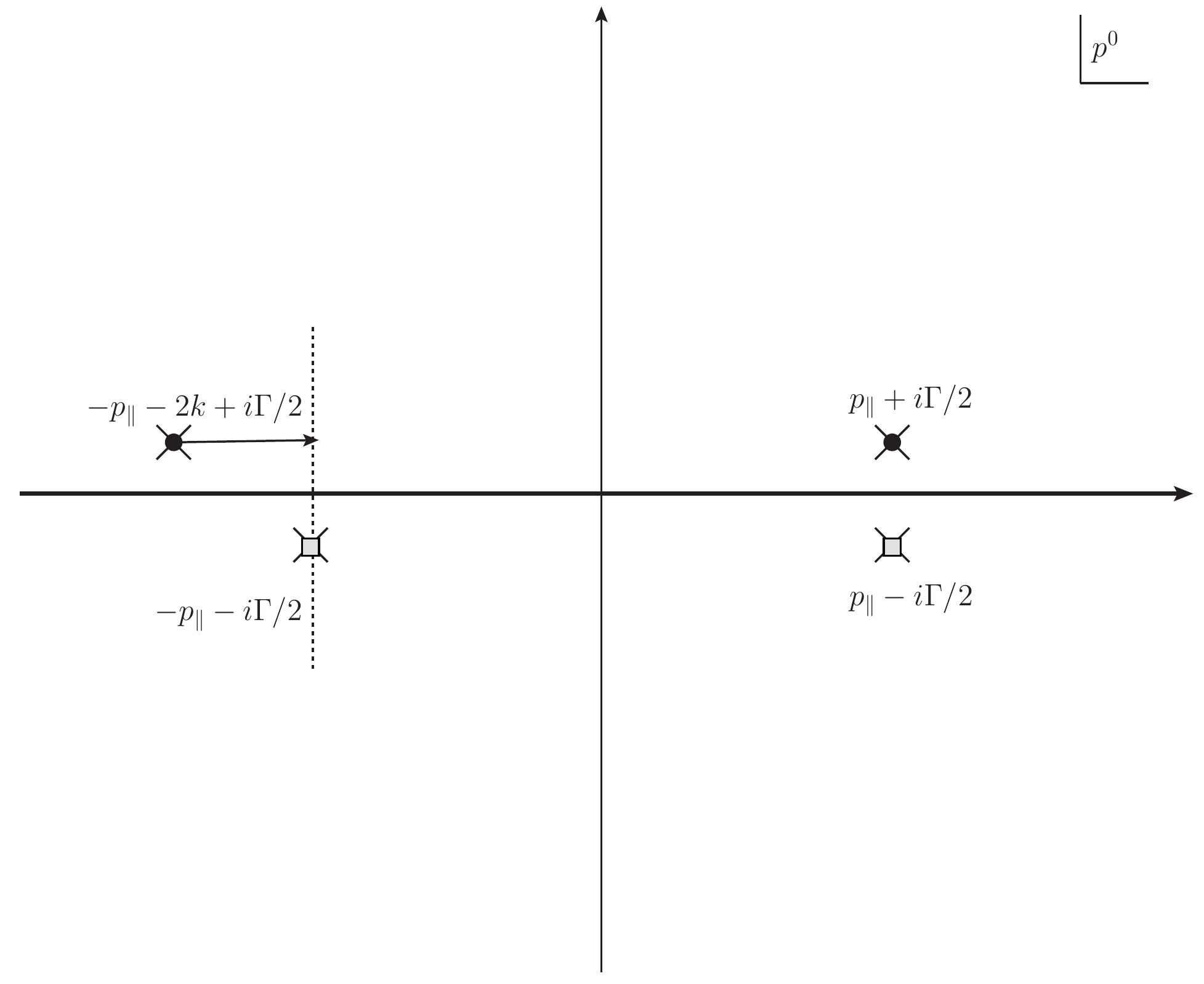}
\caption{The locations of the four poles of the frequency integrand when
$\vp$ and $\vk$ are nearly aligned. The poles above the real line belong
to $G_{\rm adv}(K+P)$ and the ones below belong to $G_{\rm ret}(P)$.
The integration contour is along the
real axis. We see that two of the poles nearly coincide at $\pp$. These
poles are said to ``pinch'' the integration contour.}
\label{fig:contour_pinch_aligned}
\end{figure}

The poles of the integrand are at the following locations: $p^0=\pm
E_\vp-\frac{i}{2}\Gamma$ and $p^0=-k^0\pm E_{\vp+\vk}+\frac{i}{2}\Gamma$.
It is straightforward to see that $E_{\bf p+k}\approx \pp + k$ when $\vp, \vk$ are $O(T)$ and collinear.
Hence the two pole positions $p^0 = E_p -i\Gamma/2$
and $p^0 = -k^0 + E_{\bf p+k} + i\Gamma/2$ almost coincide
at $\pp\equiv \vp\cdot\hk $ although they are located
on opposite halves of the contour
(c.f.~figure~\ref{fig:contour_pinch_aligned}).

We use the residue theorem to close the contour and pick up the
contributions of the poles located below the real axis. For the case where
$\pp>0$, this gives:
\begin{align}
\int\frac{dp^0}{2\pi}G_{\mathrm{Adv}}(K+P)G_{\mathrm{Ret}}(P)&\simeq\frac{-i}{2E_{\vp}(k+E_\vp-E_{\vp+\vk}-i\Gamma)(k+E_\vp+E_{\vp+\vk}-i\Gamma)}
\end{align}
When $\pp < 0$,
the factors in the denominators should be replaced by
the general expression
\begin{align}
\label{eq:pinch_difference}
E_\vp-E_{\vp+\vk}+k &\to
E_\vp\sgn(\pp)-E_{\vp+\vk}\sgn(\pp+k)+k = \delta E
\\
E_\vp+E_{\vp+\vk}+k-i\Gamma &\to
E_{\vp}\sgn(\pp)+E_{\vp+\vk}\sgn(\pp+k)+k
\simeq 2(\pp+k)
\end{align}
Therefore, the frequency integral is approximately:
\begin{align}
\int\frac{dp^0}{2\pi}G_{\mathrm{Adv}}(K+P)G_{\mathrm{Ret}}(P)
\simeq
\frac{1}{4\pp(\pp+k)(\Gamma+i\delta E)}
\end{align}
But $\Gamma$ and $\delta E$ are both of order $g^2 T$, while $\vp$ and
$\vk$ are hard. Hence, we get a $1/g^2T^3$ enhancement from the frequency
integral. These enhancements make a large class of diagrams contribute to
leading order even though we would naively expect them to be subleading.

Suppose that in the above demonstration we had not taken the integrand to
be the product of an advanced and a retarded propagator but rather $\int
G_{\mathrm{Adv}}(P)G_{\mathrm{Adv}}(P+K)$ or $\int
G_{\mathrm{Ret}}(P)G_{\mathrm{Ret}}(P+K)$. Then all poles would have been
on one side of the frequency integration contour only and we could have
closed it on the side with no poles. Since the contribution from great
circles at infinity vanishes, this means that the integral would vanish.
Hence pinching pole enhancements only arise when we have a retarded and an
advanced
propagator.
This is the reason why only $G_{aarr}$ contributes to the leading order
photon production.
Due to the fact that there is no $aa$ propagator (c.f.~Eq.(\ref{eq:Gaa}))
and that an interacting vertex must contain
an odd number of $a$-fields~\cite{Chou:1984es}, any other labelling leads
to
reduction of the number of pinching poles.

\begin{figure}[t]
\centering
\includegraphics[width=0.7\textwidth,height=3.5cm]{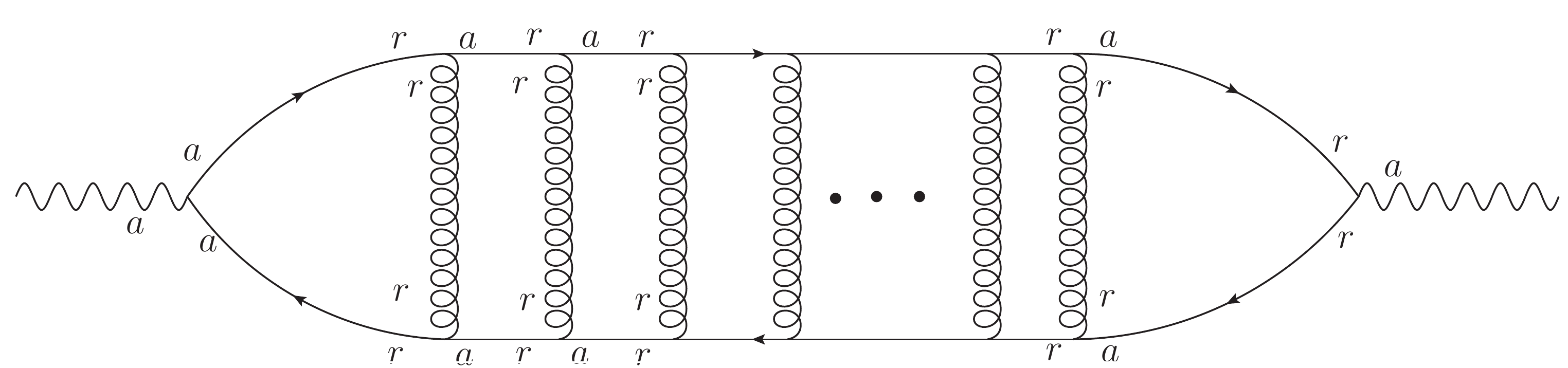}
\caption{All ladder diagrams contribute at leading order when the gluon
exchange momenta are soft.}
\label{fig:ladder_diagrams}
\end{figure}

\begin{figure}[t]
\centering
\includegraphics[width=0.7\textwidth]{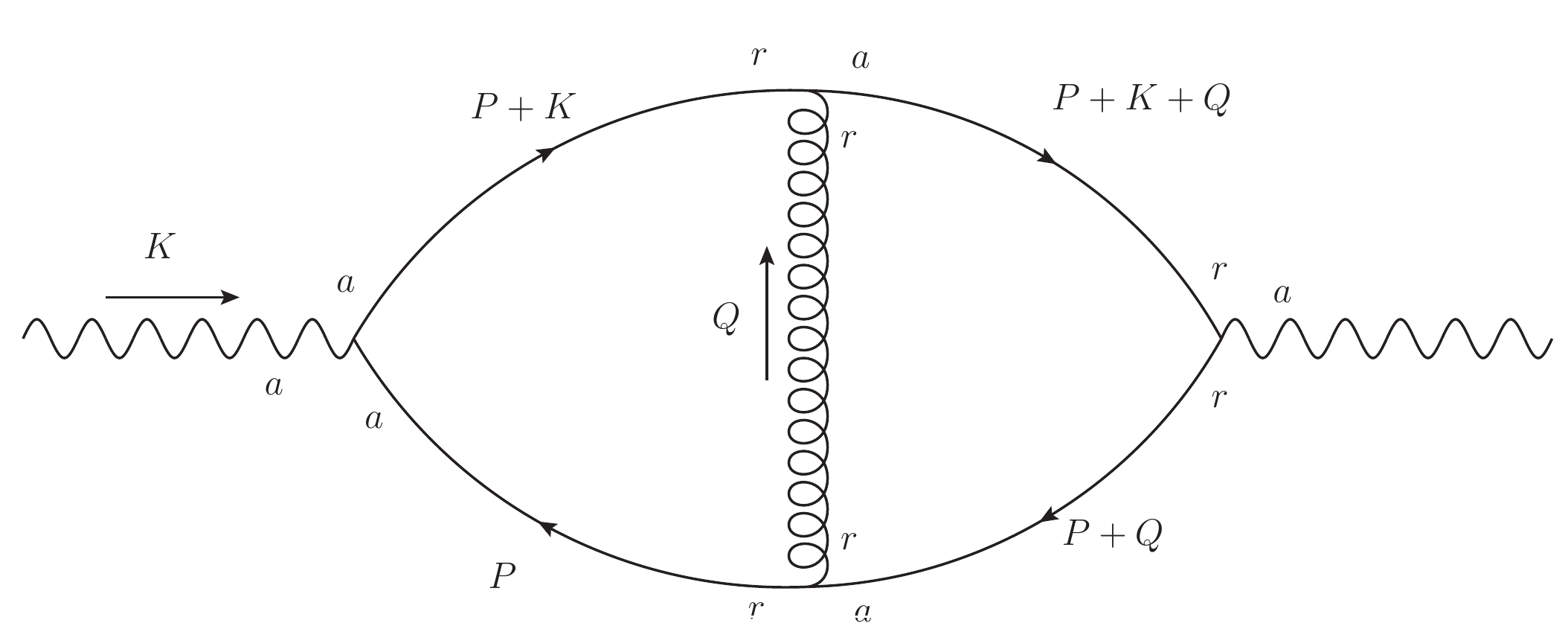}
\caption{A ladder diagram with one rung only. It has two pinching pole
enhancements and a further enhancement coming from the size of the soft
gluon $rr$ propagator.}
\label{fig:one_rung_ladder}
\end{figure}

We will illustrate general power counting arguments by estimating the size
of the ladder diagram shown in figure~\ref{fig:ladder_diagrams}.
We start with the ladder diagram that has only a single rung (figure
\ref{fig:one_rung_ladder}). Notice that the assignment of the $r,a$
indices
ensures we have a retarded and an advanced propagator in each pair of
quark
propagators.
To get a pinching pole enhancement, we need that the quark with momentum
$P$ be nearly collinear with the emitted photon. Further, if the gluon
carries a soft exchange momentum $Q\sim gT$, then it cannot disturb the
collinearity of the quark with the emitted photon. Therefore, the second
pair of propagators also gives a pinching pole enhancement. So we have a
$g^{-2}\times g^{-2}$ enhancement. A soft gluon $rr$ propagator is of
order
$g^{-3}T^{-2}$ because of the Bose-Einstein factor,
\emph{provided $\mu\leq O(T)$}. However, a soft gluon also
brings in a phase space suppression of order $g^3$ which cancels this
enhancement.

Gathering all powers, we have for suppressions:
\begin{itemize}
\item $g^2$ from the integral over $P_\perp$ which is $O(g^2)$ since this
is the near collinear regime.
\item $g^3$ from the soft $Q$ spatial integral over $Q$.
\item $g^2$ from two gluon exchange vertices.
\item $g^2$ from the $P_\perp\cdot(P+Q)_\perp$ factor arizing from the
contraction with the external photons.
\end{itemize}
and for enhancements:
\begin{itemize}
\item $g^{-2}\times g^{-2}$ from two pinching pole enhancements.
\item $g^{-3}$ from the soft gluon propagator.
\end{itemize}
Adding up all powers of $g$, we get that this two-loop
diagram is of order $g^2$, the same as the non-collinear one-loop diagram.

The above analysis easily extends to ladder diagrams with an arbitrary
number of rungs.
Each additional rung brings with it one more pinching pole
enhancement, giving a $g^{-2}$, and a soft gluon propagator of order
$g^{-3}$. It also brings in a $g^3$ suppression from the spatial integral
over the new soft gluon momentum and a $g^2$ suppression from the two
additional gauge boson exchange vertices. Adding these up, we get a net
contribution of $g^0$, so that at least all ladder diagrams contribute to
leading order.

One ought to make an important remark at this point. Since the thermal
mass
of the gluon goes roughly as $T^2+\mu^2/{2\pi^2}$, the spectral function
can be of higher order than $g^{-2}$ if $\mu$ is of order $T/g$.
In this case, the size of the soft $rr$ gluon
propagator fails to cancel the $g^3$ phase space suppression that we get
from a soft gluon loop momentum. Consequently, all ladder diagrams are
subleading when $\mu=O(T/g)$, and it becomes unnecessary to
resum the contributions of all ladder diagrams.

The rest of the power counting analysis aims at identifying which
combinations of external $r,a$ indices contribute at leading order,
showing
that all other diagram topologies are subleading, and also proving that
gauge
boson momenta of order higher or less than $gT$ also give subleading
contributions. As we have argued, as long as $\mu\ll T/g$, the size of the
propagators comprising each diagram is parametrically the same as in the
analysis of \cite{Arnold:2001ba}. We will not pursue it here in more
detail.

\section{Resummation of Ladder Diagrams}\label{sect:resummation}

In the regime where $\mu\leq O(T)$, the power counting analysis of
\cite{Arnold:2001ba} is unchanged. Therefore, the conclusion that one
needs
to resum ladder diagrams (and only these) to get the photon emissivity to
leading order is also valid. Hence, we can apply the same
resummation procedure as \cite{Arnold:2001ba}. Here we briefly outline
that resummation procedure.

\begin{figure}[t]
\centering
\includegraphics[width=0.95\textwidth]{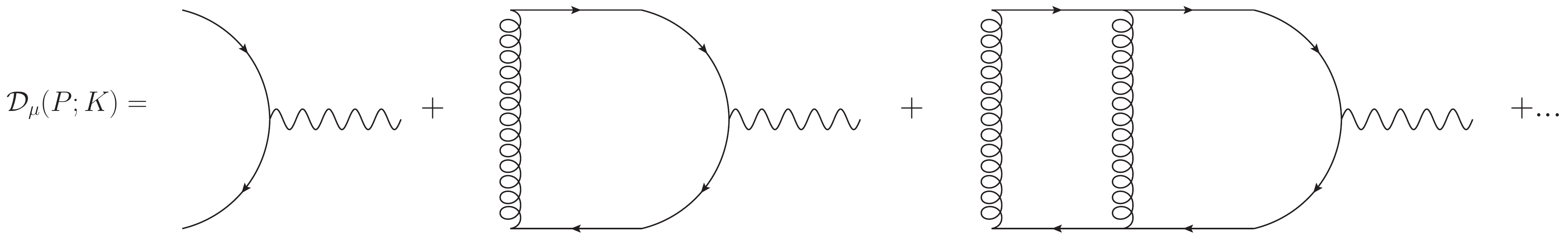}
\caption{The resummed vertex $\mathcal{D}_\mu$ sums all ladder diagrams
with one photon vertex removed. All external legs are amputated.}
\label{fig:graph_opt_D}
\end{figure}

\begin{figure}[t]
\centering
\begin{minipage}{0.3\textwidth}
\includegraphics[width=\textwidth]{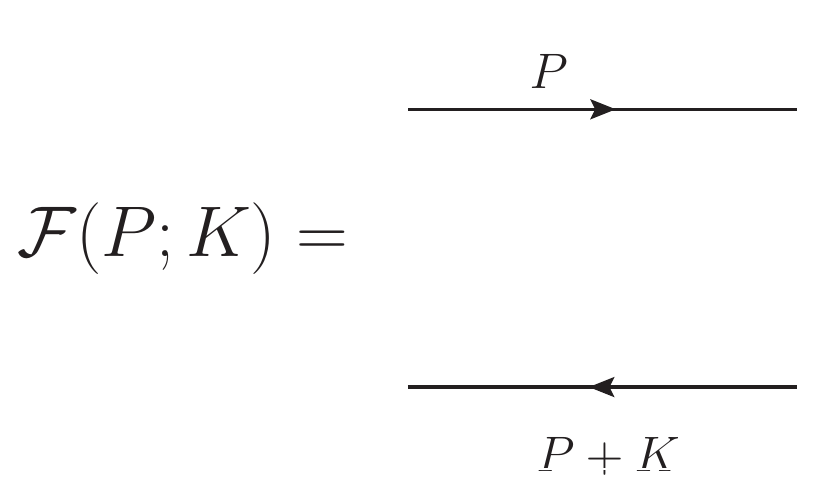}
\end{minipage}
\hspace{0.5cm}
\begin{minipage}{0.3\textwidth}
\includegraphics[width=\textwidth]{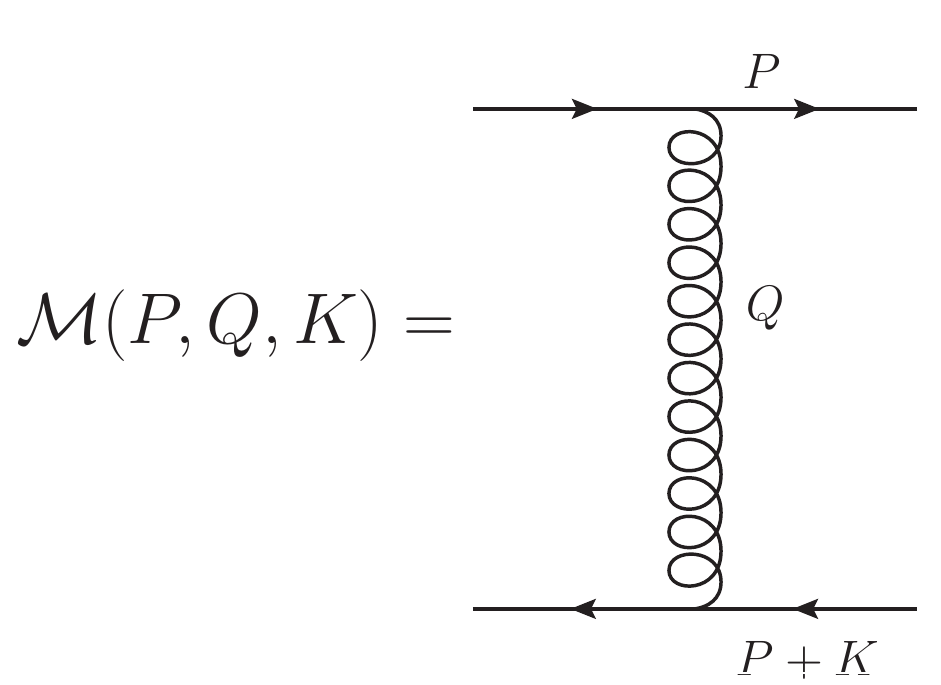}
\end{minipage}
\caption{The graphical operator $\mathcal{F}$ adds a pair of propagators
evaluated at the location of the pinching poles.
The rung $\mathcal{M}$ adds a soft gauge boson propagator.}
\label{fig:graph_opt_M}
\end{figure}

\begin{figure}[t]
\centering
\includegraphics[width=0.3\textwidth]{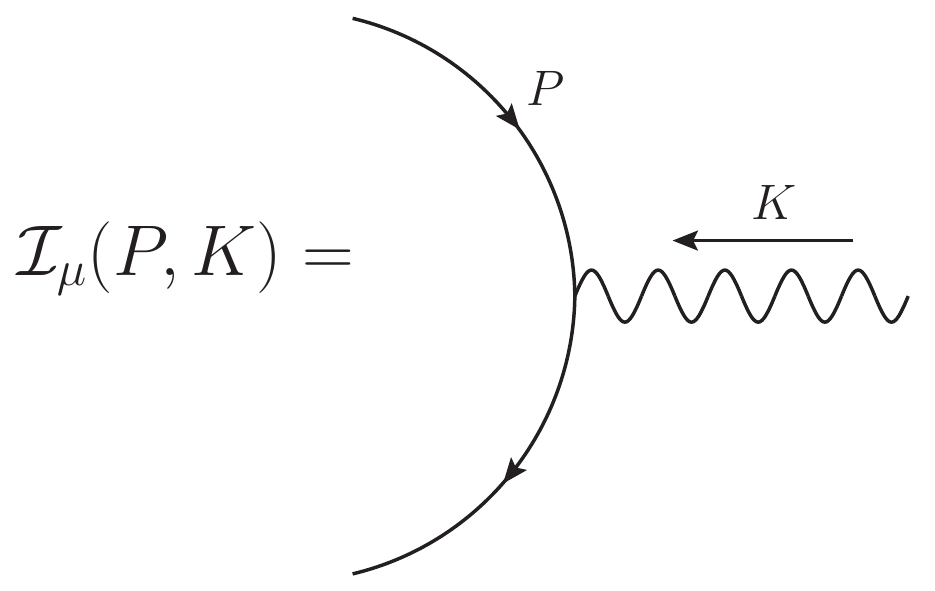}
\caption{Bare quark-photon vertex $\mathcal{I}_\mu$.}
\label{fig:graph_opt_I}
\end{figure}

Graphically, the sum we need to evaluate is illustrated in figure
\ref{fig:graph_opt_D}.
We have defined the photon resummed vertex
$\mathcal{D}_\mu$ as the sum of
all ladder diagrams with the leftmost photon vertex and pair of quark
propagators removed. From figure \ref{fig:graph_opt_D}, one sees that each
ladder is constructed from the previous one by concatenating a gluon rung
attached to two quark-gluon vertices and a pair of quark propagators to
the
previous one, where the quark propagators must be evaluated at the
location
of pinching poles. The ladder diagrams form a geometric series.
\begin{align}
\label{eq:geo_series}
\mathcal{D}_\mu
=\mathcal{I}_\mu
+\mathcal{M}\mathcal{F}\mathcal{I}_\mu
+\mathcal{M}\mathcal{F}\mathcal{M}\mathcal{F}\mathcal{I}_\mu
+\mathcal{M}\mathcal{F}\mathcal{M}\mathcal{F}
\mathcal{M}\mathcal{F}\mathcal{I}_\mu+...
\end{align}
The graphical operator
$\mathcal{F}$ adds the pinching pole pair of propagators. In the above
equation, each ${\cal F}$ contributes a pair of pinching poles:
\begin{align}
\label{eq:Fdef}
\mathcal{F}(P;K)&=
G_{\rm Adv}(K+P)G_{\rm Ret}(P)
\nonumber\\
&\approx
\frac{-1}{4\pp(\pp+k)}\frac{1}{\Gamma+i\delta
E}4\pi\delta\left[2p^0+k^0-E_{\vp}\sgn(\pp)-E_{\vp+\vk}\sgn(\pp+k)\right]
\end{align}
%
%
The operator $\mathcal{M}$ adds the rung
as illustrated in figure~\ref{fig:graph_opt_M}
\begin{align}
\mathcal{M}(P,Q,K)
=ig^2C_{F}4\pp(\pp+k)\hat{K}_{\mu}\hat{K}_{\nu}G_{rr}^{\mu\nu}(Q)
\end{align}
Finally, $\mathcal{I}_\mu$ is the bare photon vertex on the far end of the
ladder diagram as illustrated in figure \ref{fig:graph_opt_I}.

The resummed vertex ${\cal D}_\mu$ then satisfies the following
integral equation:
\begin{align}
\label{eq:Int}
\mathcal{D}_\mu(P,K)&=\mathcal{I}_\mu(P,K)
+\int_{Q}\mathcal{M}(P,Q,K)\mathcal{F}(P+Q,K)\mathcal{D}_\mu(P+Q,K)
\end{align}
After defining the variable $f^{\mu}(\vp,\vk)\equiv 4
\pp(\pp+k)\int\frac{dp^0}{2\pi}\mathcal{F}(P,K)\mathcal{D}^{\mu}(P,K)$,
the
integral equation (\ref{eq:Int}) becomes \cite{Arnold:2001ba}:
\begin{align}
\label{eq:Intt}
2P^\mu+K^\mu &= i\delta E f^{\mu}(\vp,\vk)
+\int \frac{d^3 \vq}{(2\pi)^3}\mathcal{C}(\vq)
\left[f^\mu(\vp,\vk)-f^\mu(\vp+\vq_{\perp},\vk)\right]
\end{align}
After applying a sum rule, the collision kernel $\mathcal{C}(\vq)$ is
found to have the simple form \cite{Aurenche:2002pd}
\begin{align}
\int {dq_{\|}\over 2\pi}\,
{\cal C}({\bf q})
=\frac{1}{\vq_\perp^2} -\frac{1}{\vq_\perp^2+m_D^2}
\end{align}
with ${\bf p}$ and ${\bf k}$ strictly colinear.
The term $\delta E$ is the difference between the locations of
the pinching poles of the quark propagators, as seen in
Eq.(\ref{eq:pinch_difference}). Through the energies of the quarks,
it includes their asymptotic thermal masses $m_{\infty}$.
Therefore, the chemical potential enters through both $m_D$ and
$m_{\infty}$.
Recall that
\begin{align}
m_D^2&=g^2\left(\frac{N_f T^2}{6}+\frac{N_c T^2}{3}
+\frac{N_f\mu^2}{2\pi^2}\right)\\
m_\infty^2&=\frac{g^2 C_{2}(F)}{4}\left(T^2+\frac{\mu^2}{\pi^2}\right)
\end{align}

Finally, the contribution of the $2\mapsto 3$ and $3\mapsto 2$ processes
to
the Wightman correlator (and thus to the photon emissivity) is given by:
\begin{align}
\label{eq:Wightman}
W_{LPM}^{\mu\nu}(K)&=\int\frac{d^4P}{(2\pi)^4}\frac{(\pp+k)^2+\pp^2}{\pp(\pp+k)}\;n_f(p^0+k^0-\mu)[1-n_f(p^0-\mu)]\times\nonumber\\
&\quad\times\mathcal{I}^{\mu}(P,K)\R[\mathcal{F}(P,K)\mathcal{D}^\nu(P,K)]
\end{align}

\section{Numerical Calculations}\label{sect:numerical_results}

Following a method developed by Aurenche et al.~\cite{Aurenche:2002wq},
we can solve equation \eqref{eq:Int} numerically by solving the
differential
equation in impact parameter space.
As in Ref.\cite{Arnold:2001ms}, we decompose the photon emission rate
as follows
\begin{align}
(2\pi)^3\frac{d\Gamma}{d^3\vk}
&=
{\cal A}(k)
\left[
\ln(T/m_\infty) + {1\over 2}\ln(2k/T)
+
C_{2\leftrightarrow 2}(k/T) + C_{\rm brem+ann}(k/T)
\right]
\end{align}
where
\begin{align}
{\cal A}(k) = 2\alpha_{\rm EM}\left( d_F\sum_{r=u,d,s} {q_r}^2 \right)
{m_\infty^2\over k} n_f(k)
\end{align}
Here, $k$ is the magnitude of the emitted photon's momentum and
$n_f$ is the Fermi-Dirac factor.

\begin{figure}
\centering
\includegraphics[width=0.75\textwidth]{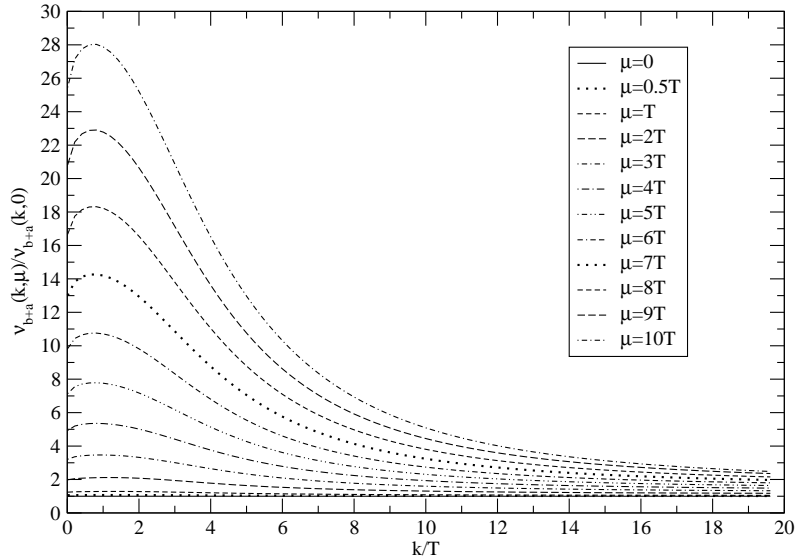}
\caption{
Plots of the ratios $\nu_{b+a}(\mu)/\nu_{b+a}(\mu=0)$ for $0
\le \mu \le 10T$.
As $\mu/T$ increases, so does the ratio.
}
\label{fig:photon_plot}
\end{figure}

Since the $2\leftrightarrow 2$ part of the spectrum has been already
calculated \cite{Dumitru:1993us,Traxler:1994hy,Dutta:1999dy,He:2005yb},
we only plot $\nu_{b+a}(k) \equiv {\cal A}(k)C_{\rm brem+ann}(k/T)$
in figure \ref{fig:photon_plot}.
The plots show the contribution of the $2\mapsto 3$ and $3\mapsto 2$
processes with the full treatment of the LPM effect. The temperature of
the
plasma was taken to be $250 \mathrm{MeV}$.
%
%
\begin{figure}[t]
\centering
\includegraphics[width=0.75\textwidth]{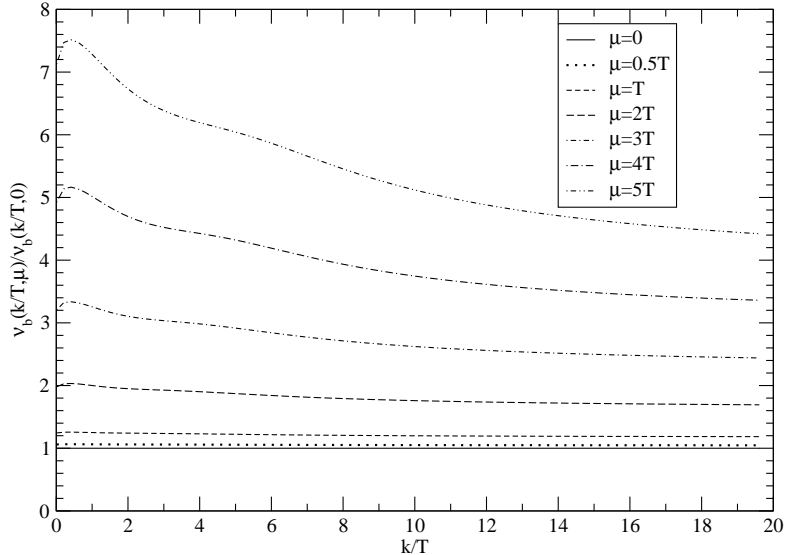}
\caption{
Combined $q$ and $\bar q$
bremsstrahlung contribution to the photon emission.
The ratio against the $\mu=0$ case {\em increases} with increasing $\mu/T$.
}
\label{fig:brem}
\end{figure}
\begin{figure}[t]
\centering
\includegraphics[width=0.75\textwidth]{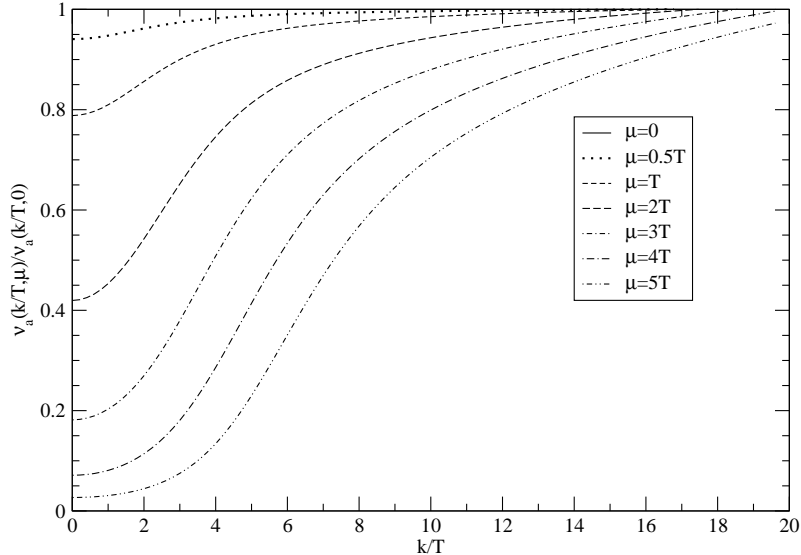}
\caption{
Pair annihilation contribution.
The ratio against the $\mu=0$ case {\em decreases} with increasing $\mu/T$.
}
\label{fig:annih}
\end{figure}
We see from figure \ref{fig:photon_plot} that as the ratio $\mu/T$
increases,
the photon production rate also increases.
This finding is consistent with the behavior of the $2\leftrightarrow 2$
emission rates.

One of the reasons
for this enhancement turned out to be just the statistical
factors in Eq.(\ref{eq:Wightman}).
These factors represent 3 different processes depending on the sign
of $p^0$ and the relative sizes of $|p^0|$ and $k^0 = k > 0$.
To illustrate the effect of non-zero $\mu$ in each of the three processes,
consider the following ratio
of the statistical factors in Eq.(\ref{eq:Wightman}) and the same statistical factors
with $\mu=0$.
\begin{align}
r &= {n_f(p^0+k^0-\mu)(1-n_f(p^0-\mu))
\over n_f(p^0+k^0)(1-n_f(p^0))}
\end{align}
When $p^0 > 0$, the underlying physical process is the
bremsstrahlung from the quarks. In this case, the ratio is mostly greater
than 1. Hence the rate is enhanced.
This reflects the fact that a positive chemical potential enhances the
number of quarks more than the Pauli-blocking factor reduces the emission rate.
When $p^0 < 0$, $(1-n_f(p^0-\mu))=n_f(|p^0|+\mu)$ becomes the anti-quark
phase space density. For $p^0 < 0$ and $k < |p^0|$,
the factor $n_f(-|p^0|+k-\mu) = 1-n_f(|p^0|-k+\mu)$
represents the Pauli-blocking factor for the anti-quark bremsstrahlung.
In this case, a positive chemical potential reduces the phase density of
anti-quarks but enhances the Pauli-blocking factor. But since
Pauli-blocking
can never be enhanced above 1, the effect is to reduce the photon
emission rate.
The combined effect of quark bremsstrahlung and anti-quark bremsstrahlung
is
a net enhancement as shown in
figure~\ref{fig:brem}.

For $p^0 < 0$ and $k > |p^0|$,
the factor $n_f(-|p^0|+k-\mu)$ represents the density of quarks that
can annihilate with anti-quarks to produce a photon with energy $k$.
In this case $r$ is always smaller than $1$,
reflecting the fact that the annihilation process is necessarily
controlled by the lesser number of anti-quarks.
In figure~\ref{fig:annih}, we show the
contribution
of the annihilation process to the photon emission rate for various
values of $\mu$.

\begin{figure}[t]
\centering
\includegraphics[width=0.75\textwidth]{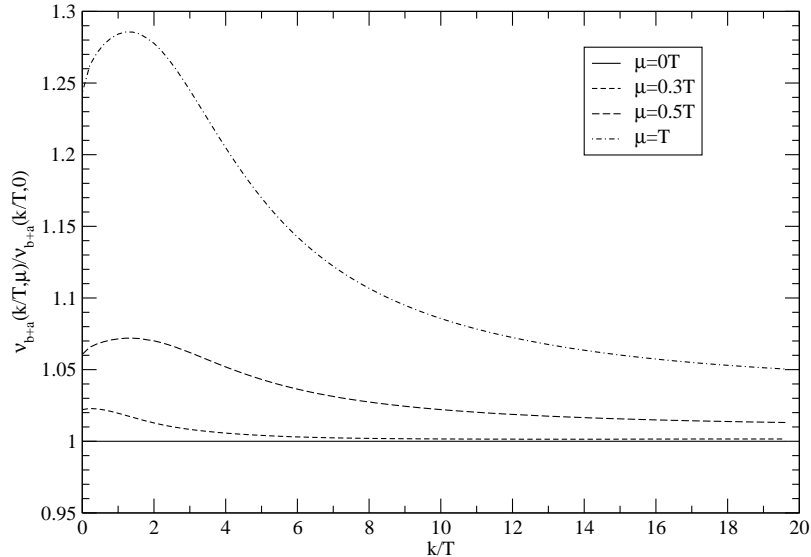}
\caption{
For $\mu$ of the order of the temperature $T$ of the plasma,
the
photon emission rate increases by about $10\%$ relative to the $\mu=0$
value.
}
\label{fig:photon_plot_restricted}
\end{figure}

Among these three effects, the enhancement of the quark-bremsstrahlung
dominates because it increases much faster than the reductions in the
annihilation contribution and the anti-quark bremsstrahlung. Therefore,
overall, the effect of having $\mu> 0$ is to enhance the photon production
at the same temperature as shown in figures~\ref{fig:photon_plot_restricted} and \ref{fig:photon_plot}.
This trend was also observed in studies of $2\leftrightarrow 2$
processes \cite{Traxler:1994hy}.
In one previous study of $2 \leftrightarrow 2$ processes
\cite{Dumitru:1993us},
it was found that increasing the chemical potential {\em decreases} photon
production.
However, this was for constant {\em energy density} instead of constant
temperature.

%
%
Phenomenologically, the baryon chemical potential at RHIC is about 30\,MeV
and at SPS it is about 240\,MeV\cite{BraunMunzinger:2001as}.
The quark chemical potential is therefore about $10\,\hbox{MeV}$
and $80\,\hbox{MeV}$ for RHIC
and SPS, respectively.
With $T = O(200\,\hbox{MeV})$, the enhancement at RHIC is negligibly small
and it will be modest at SPS, no more than 5\,\%.


\section{Conclusion}\label{sect:conclusion}

In this paper, we have explored the effect of non-zero baryon chemical
potential in thermal photon production.
Following the zero chemical potential study \cite{Arnold:2001ba},
we have computed photon production from non-Abelian plasmas at leading
order
by resumming ladder diagrams to fully incorporate the LPM effect.
After a careful analysis,
we have found that as long as $\mu \leq O(T)$, the formulation in
\cite{Arnold:2001ba} is still valid with appropriate changes in the
statistical factors, thermal masses and the Debye mass.
However, when $\mu = O(T/g)$, resummation of ladder diagrams is no longer
necessary because thermal masses become $O(T)$ instead of $O(gT)$.
Hence, inverse powers of thermal masses no longer provide enhancements.

Numerically, it is found that the inclusion of a chemical potential up to
$\mu \simeq T$ enhances the photon emission rate moderately, up to $30\,\%$
when $\mu = T$.
This trend is also valid for hard photons from the $2\leftrightarrow 2$
process\cite{Traxler:1994hy}.
Since the quark chemical potential is relatively small compared to
typical QGP temperatures at RHIC and the LHC,
we expect a relatively small effect from a finite
$\mu$ on thermal photon production at RHIC and the LHC although it could
be significant at SPS energies and also for the lower energy runs at RHIC.

\section*{Acknowledgement}
H.G and S.J.~are supported in part by the Natural Sciences and
Engineering Research Council of Canada.
H.G is also supported in part by le Fonds Nature et Technologies of
Qu\'ebec.
We gratefully acknowledge
many discussions with C.Gale, G.D.Moore and J.Cline. H.G.~also would like
to thank J.F.~Paquet for his help in numerical calculations.

\appendix

\section{KMS Condition with Finite $\mu$}

In this appendix, our goal is to prove equation (\ref{eq:change_basis}).
This equation was proven in reference \cite{Wang:1998wg} for the case of
scalar fields.
Their argument exploits the KMS condition to derive a system of equations
that allows them
to solve for $G_{rrrr}$, $G_{rrra}$, $G_{rrar}$, $G_{rarr}$, $G_{arrr}$,
$G_{raar}$,
$G_{rara}$, and $G_{rraa}$ in terms of $G_{aarr}$, $G_{aaar}$,
$G_{aara}$, $G_{araa}$, $G_{raaa}$, $G_{arra}$, $G_{arar}$ and their
complex conjugates. Their procedure also goes through for Dirac fermions in
the grand
canonical ensemble provided we modify the usual KMS condition to include a
chemical potential.

Consider first the commutation relation between Dirac field operators and
the conserved charge $\hat{Q}$:
\begin{align}
[\hat{Q},\psi]=-\psi &\Rightarrow
e^{\beta\mu\hat{Q}}\psi=e^{-\beta\mu}\psi e^{\beta\mu\hat{Q}}\nonumber\\
\label{eq:commutator_Q_psi}[\hat{Q},\bpsi]=\bpsi &\Rightarrow
e^{\beta\mu\hat{Q}}\bpsi=e^{\beta\mu}\bpsi e^{\beta\mu\hat{Q}}
\end{align}
For Green functions in the $1,2$ basis, these commutation relations allow
us to relate Green functions whose $1,2$ index assignments are opposite
each other. Indeed, consider:
\begin{align}
G_{2..21..1}(x_1,...,x_n)&=\langle
\tilde{T}(\dg{\psi}(x_1)...\dg{\psi}(x_r)\psi(x_{r+1})...\psi(x_m))\times\nonumber\\
&\quad\times
T(\dg{\psi}(x_{m+1})...\dg{\psi}(x_l)\psi(x_{l+1})...\psi(x_n))\rangle_{\beta,\mu}
\end{align}
Note that the $2$ index refers to the space-time points $x_1,...,x_m$ and
the $1$ index refers to the points $x_{m+1},...,x_{n}$. We have also
omitted spinor indices in the above for simplicity. Using the fact
that $e^{\beta \hH}$ is a time evolution operator in imaginary time
($\psi(t+i\beta)=e^{-\beta\hH}\psi(t)e^{\beta\hH}$), we obtain:
\begin{align}
&G_{2..21..1}(x_1,...,x_n)=\nonumber\\
&\quad\langle
\tilde{T}(e^{\beta\hH}\dg{\psi}(x_1^0+i\beta)...\dg{\psi}(x_r^0+i\beta)\psi(x_{r+1}^0+i\beta)...\psi(x_m^0+i\beta)e^{-\beta\hH})\times\nonumber\\
&\quad \times
T(\dg{\psi}(x_{m+1})...\dg{\psi}(x_l)\psi(x_{l+1})...\psi(x_n))\rangle_{\beta,\mu}
\end{align}
The operators $e^{\pm\beta
\hH}$ can be pulled outside of the reversed time-ordering symbol.
Therefore,
when writing out the thermal average explicitly in terms of the density
operator $e^{-\beta(\hH-\mu\hat{Q})}$ , we obtain:
\begin{align}
&G_{2..21..1}(x_1,...,x_n)=\nonumber\\
&\quad\frac{1}{Z}\Tr(e^{\beta\mu\hat{Q}}
\tilde{T}(\dg{\psi}(x_1^0+i\beta)...\dg{\psi}(x_r^0+i\beta)\psi(x_{r+1}^0+i\beta)...\psi(x_m^0+i\beta))\times\nonumber\\
&\quad \times
e^{-\beta\hH}T(\dg{\psi}(x_{m+1})...\dg{\psi}(x_l)\psi(x_{l+1})...\psi(x_n)))
\end{align}

We want to commute the operators inside the reverse time-ordering operator
past $e^{\beta\mu\hat{Q}}$ and use the cyclicity of the trace to take them
to the right of the operators inside the time-ordering symbol. Equation
\eqref{eq:commutator_Q_psi} shows that in doing this, we pick up a factor
of $e^{\beta\mu}$ when we are commuting a $\dg{\psi}$ operator and a
factor
of $e^{-\beta\mu}$ for a $\psi$ operator. Let us define the symbol
$\sigma_{i}$ to be equal to $1$ if the space-time point $x_i$ has a
$\dg{\psi}$ operator insertion at it, or $-1$ if it has a $\psi$
insertion.
With this notation, we obtain the following generalization of the KMS
boundary condition to $n$-point functions:
\begin{align}
&G_{2..21..1}(x_1,...,x_n)=\nonumber\\
&\quad\frac{1}{Z}e^{\sum_{i|a_i=2}\sigma_{i}\beta\mu}\Tr(e^{-\beta(\hH-\mu\hat{Q})}T(\dg{\psi}(x_{m+1})...\dg{\psi}(x_l)\psi(x_{l+1})...\psi(x_n)) \times\nonumber\\
&\quad \times
\tilde{T}(\dg{\psi}(x_1^0+i\beta)...\dg{\psi}(x_r^0+i\beta)\psi(x_{r+1}^0+i\beta)...\psi(x_m^0+i\beta)))
\end{align}
Finally, we use the well-known formula $\phi(t+a)=e^{a\partial_t}\phi(t)$
to obtain:
\begin{align}
&G_{2..21..1}(x_1,...,x_n)=\nonumber\\
&\quad e^{\beta\sum_{i|a_i=2}(i\partial_{t_i}+\sigma_{i}\mu)}\langle
T(\dg{\psi}(x_{m+1})...\dg{\psi}(x_l)\psi(x_{l+1})...\psi(x_n))
\times\nonumber\\
&\quad \times
\tilde{T}(\dg{\psi}(x_1^0)...\dg{\psi}(x_r^0)\psi(x_{r+1}^0)...\psi(x_m^0))\rangle_{\beta,\mu}
\end{align}

We are allowed to take the $e^{\sum_{i|a_i=2}i\beta\partial_{t_i}}$
operator outside of the time-ordering symbol provided we take
time-ordering
to be given by the $T^{*}$ prescription. That is, the time-ordered
$n$-point function is \emph{defined} to be the path integral of the
product
of the $n$ operators weighted by the exponential of $i$ times the action.
For a discussion of this prescription, see for example
\cite{Itzykson:1980rh}.

We define the ``tilde conjugate'' $\tilde{G}$ of a Green function $G$ to
have same $1,2$ index assignment but with the time-ordering symbols
reversed. Therefore, we can express our previous result as:
\begin{align}
G_{2..21..1}(x_1,...,x_n)=e^{\beta\sum_{i|a_i=2}(i\partial_{t_i}+\sigma_{i}\mu)}\tilde{G}_{1..12..2}(x_1,...,x_n)\label{eq:KMS_generalized}
\end{align}

In \cite{Wang:1998wg}, Wang and Heinz use the fact that in momentum space,
tilde conjugation is equivalent to complex conjugation. As this is a key
fact, we proceed to extend it to fermion fields at finite chemical
potential. For notational simplicity, we will focus on the case where there
are four fermion operators two of which bear ``$1$'' indices, and the other
two of which bear ``$2$'' indices. Although this is not necessary for this
discussion, we have also included the $\gamma$ matrices which appear in the
current-current correlator \ref{eq:current_current}. Extensions to other
cases should be obvious.
\begin{align}
G_{2211}(Q_1,Q_2,Q_3,Q_4)&=\int\prod_{i=1}^{4}d^{4}x_i\;e^{iQ_i\cdot
x_i}\langle\tilde{T}(\bpsi(x_1)\gamma^{\mu}\psi(x_2))T(\bpsi(x_3)\gamma^{\nu}\psi(x_4))\rangle_{\beta,\mu}
\end{align}
Taking a complex conjugation and changing $x_i \to -x_i$, we obtain:
\begin{align}
G_{2211}^{*}(Q_1,Q_2,Q_3,Q_4)&=\int\prod_{i=1}^{4}d^{4}x_i\;e^{iQ_i\cdot
x_i}\langle
T(\bpsi(-x_4)\gamma^{\nu}\psi(-x_3))\tilde{T}(\bpsi(-x_2)\gamma^{\mu}\psi(-x_1))\rangle_{\beta,\mu}
\end{align}
Next, we make use of $C\!PT$ invariance. Recall that the Dirac bilinear
$\bpsi \gamma^{\mu} \psi$
transforms as follows under the anti-unitary $C\!PT$ transformation
$\Theta$:
\begin{align}
\Theta \bpsi(x)\gamma^{\mu}\psi(y)\Theta^{-1}
&=-\bpsi(-y)\gamma^{\mu}\psi(-x)
\end{align}
Then, using anti-unitarity, thermal expectations of a product of fields can
be related as follows:
\begin{align}
\langle
&T(\bpsi(-x_4)\gamma^{\nu}\psi(-x_3))\tilde{T}(\bpsi(-x_2)\gamma^{\mu}\psi(-x_1))\rangle_{\beta,\mu}^*
\nonumber\\
&=\frac{1}{Z}\sum_{n}\bra{n}e^{-\beta(\hH-\mu\hat{Q})}T(\bpsi(-x_4)\gamma^{\nu}\psi(-x_3))
\tilde{T}(\bpsi(-x_2)\gamma^{\mu}\psi(-x_1))\ket{n}^*
\nonumber\\
&=\frac{1}{Z}\sum_{n}
\bra{\Theta n}
\Theta
e^{-\beta(\hH-\mu\hat{Q})}T(\bpsi(-x_4)\gamma^{\nu}\psi(-x_3))\tilde{T}(\bpsi(-x_2)\gamma^{\mu}\psi(-x_1))
\Theta^{-1}\ket{\Theta n}
\nonumber\\
&=\frac{1}{Z}\sum_{n^\prime}\bra{n^\prime} e^{-\beta(\hH+\mu\hat{Q})}
T(\Theta\bpsi(-x_4)\gamma^{\nu}\psi(-x_3)\Theta^{-1})
\tilde{T}(\Theta\bpsi(-x_2)\gamma^{\mu}\psi(-x_1)\Theta^{-1})\ket{n^\prime}\nonumber\\
&=\frac{1}{Z}\sum_{n^\prime}\bra{n^\prime}
e^{-\beta(\hH+\mu\hat{Q})}T(\bpsi(x_3)\gamma^{\nu}\psi(x_4))\tilde{T}(\bpsi(x_1)\gamma^{\mu}\psi(x_2))\ket{n^\prime}\nonumber\\
&=\langle
T(\bpsi(x_3)\gamma^{\nu}\psi(x_4))\tilde{T}(\bpsi(x_1)\gamma^{\mu}\psi(x_2))\rangle_{\beta,-\mu}
\end{align}
We have used $C\!PT$ invariance in commuting $\Theta$ past $e^{-\beta\hH}$
and relabeling the eigenstates $\ket{n^\prime}=\Theta\ket{n}$. Further,
note that the sign of $\mu$ changes as we commute
$\Theta$ past $e^{\beta\mu\hat{Q}}$. Combining the above with our previous
result, we obtain:
\begin{align}
G_{2211}^{*}(Q_1,Q_2,Q_3,Q_4)&=\int\prod_{i=1}^{4}d^{4}x_i\;e^{iQ_i\cdot
x_i}\langle
T(\bpsi(-x_4)\gamma^{\nu}\psi(-x_3))\tilde{T}(\bpsi(-x_2)\gamma^{\mu}\psi(-x_1))
\rangle_{\beta,\mu}\nonumber\\
&=
\int\prod_{i=1}^{4}d^{4}x_i\;e^{iQ_i\cdot x_i}
\langle
T(\bpsi(x_2)\gamma^{\mu}\psi(x_1))
\tilde{T}(\bpsi(x_4)\gamma^{\nu}\psi(x_3))
\rangle_{\beta,-\mu}\nonumber\\
&=\tilde{\bar{G}}_{2211}(Q_1,Q_2,Q_3,Q_4)
\end{align}
where $\bar{G}$ denotes the charge conjugate ($\bpsi\leftrightarrow \psi$)
of $G$.
Recapitulating, we have the two equations:
\begin{align}
G_{2..21..1}(K_1,...,K_n)&=e^{\beta\sum_{i|a_i=2}(k_{i}^{0}+\sigma_{i}\mu)}\tilde{G}_{2..21..1}^{*}(K_1,...,K_n)\nonumber\\
\label{eq:final_wang_heinz}\tilde{G}_{2..21..1}(K_1,...,K_n)&=\bar{G}^{*}_{2..21..1}(K_1,...,K_n)
\end{align}
The above can be taken as the starting point to reproduce the derivation of
Heinz and Wang. However, the latter also uses the energy conservation
condition $\sum_{i}k_i^0=0$ and the identity
$n_{b}(k_i^0)+n_{b}(-k_i^0)=-1$. In order for the derivation of Heinz and
Wang to proceed in the same manner, we need suitable generalizations of
these to treat fermions at finite chemical potential. By charge
conservation, we must have that
$\sum_{i=1}^{n}\sigma_i=0$ -- In other words, the Green function in
question must have an equal number of $\dg{\psi}$ and $\psi$ insertions to
not vanish. Therefore, we can define $\tilde{k}_i^0\equiv
k_{i}^{0}+\sigma_{i}\mu$ and maintain ``energy conservation''
$\sum_{i=1}^{n}\tilde{k}_i^0=0$. Also, by performing the customary
replacement $n_{b}(k_i^0)\mapsto -n_{f}(\tilde{k}_i^0)$, we preserve the
relation $n_{b}(k_i^0)+n_{b}(-k_i^0)=-1$ since $n_f(E)+n_f(-E)=1$ in
general.

 The only relations among the $k_i^0$'s that Heinz and Wang use is energy
conservation, and the only property of the Bose-Einstein distributions
they
use is $n_{b}(k_i^0)+n_{b}(-k_i^0)=-1$. This is because after they have
used the KMS condition like we have done above, their work consists
of algebraic manipulations such as solving large systems
of equations involving the Bose-Einstein distribution. In their work, it is
possible to treat $G$
and $G^*$ as independent variables rather than considering the real and
imaginary parts of $G$.
Hence, it causes no harm to their derivation to replace $G^*$ by
$\bar{G}^*$ as is required by \eqref{eq:final_wang_heinz}.
We conclude that we could in
principle perform the exact same manipulations as Heinz and Wang by
relabeling every $n_b(k_i^0)$ as $-n_f(\tilde{k}_i^0)$. The net result is
that we make the same replacements in their final results, bearing in mind
that $\dg{\psi}$ insertions come with a $-n_f(k^0+\mu)$ and $\psi$
insertions come with $-n_f(k^0-\mu)$.

In reference \cite{Wang:1998wg}, Heinz and Wang obtain
\begin{align}
\alpha_1= n_b(q_1^0)n_b(q_2^0)
\end{align}
and
\begin{align}
\beta_1=-(1+n_b(q_3^0))(1+n_b(q_4^0))\frac{1+n_b(q_1^0)+n_b(q_2^0)}{1+n_b(q_3^0)+n_b(q_4^0)}
\end{align}
By making the replacements we have prescribed above, we get:
\begin{align}
\alpha_1= n_f(q_1^0+\mu)n_f(q_2^0-\mu)
\end{align}
and
\begin{align}
\beta_1=-(1-n_f(q_3^0+\mu))(1-n_f(q_4^0-\mu))\frac{1-n_f(q_1^0+\mu)
-n_f(q_2^0-\mu)}{1-n_f(q_3^0+\mu)-n_f(q_4^0-\mu)}
\end{align}
as required.

\section{A Simple Identity Involving Charge Conjugates}

In this appendix, we want to show that
$\bar{G}_{aarr}(K_1,K_2,K_3,K_4)=G_{aarr}(K_1,K_2,K_3,K_4)$. It is
sufficient to prove the corresponding identity in position space.
\begin{align*}
G_{aarr}(x_1,x_2,x_3,x_4)&\equiv \langle
T_{C}(\bpsi_a(x_1)\psi_a(x_2)\bpsi_r(x_3)\psi_r(x_4))\rangle_{\beta,\mu}\\
\Rightarrow \bar{G}_{aarr}(x_1,x_2,x_3,x_4)&= \langle
T_{C}(\bpsi_a(x_2)\psi_a(x_1)\bpsi_r(x_4)\psi_r(x_3))\rangle_{\beta,-\mu}\\
&=\frac{1}{Z}\sum_{n}\bra{n}e^{-\beta(\hH+\mu\hat{Q})}T_{C}
(\bpsi_a(x_2)\psi_a(x_1)\bpsi_r(x_4)\psi_r(x_3))\ket{n}\\
&=\frac{1}{Z}\sum_{n}\bra{Cn}Ce^{-\beta(\hH+\mu\hat{Q})}T_{C}
(\bpsi_a(x_2)\psi_a(x_1)\bpsi_r(x_4)\psi_r(x_3))\dg{C}\ket{Cn}\\
&=\frac{1}{Z}\sum_{n^\prime}\bra{n^\prime}e^{-\beta(\hH-\mu\hat{Q})}CT_{C}
(\bpsi_a(x_2)\psi_a(x_1)\bpsi_r(x_4)\psi_r(x_3))\dg{C}\ket{n^\prime}\\
&=\frac{1}{Z}\sum_{n^\prime}\bra{n^\prime}e^{-\beta(\hH-\mu\hat{Q})}T_{C}
(C\bpsi_a(x_2)\psi_a(x_1)\dg{C}C\bpsi_r(x_4)\psi_r(x_3)\dg{C})\ket{n^\prime}\\
&=\frac{1}{Z}\sum_{n^\prime}\bra{n^\prime}e^{-\beta(\hH-\mu\hat{Q})}T_{C}
(\bpsi_a(x_1)\psi_a(x_2)\bpsi_r(x_3)\psi_r(x_4))\ket{n^\prime}\\
&=G_{aarr}(x_1,x_2,x_3,x_4)
\end{align*}

\section{Computation of Self-Energies at Finite Chemical Potentials}

In this appendix, we closely follow
the zero $\mu$ analysis in \cite{Blaizot:2001nr}
to compute self-energies in the hard thermal loop approximation with a
finite
chemical potential.

\begin{figure}[t]
\centering
\includegraphics[width=0.5\textwidth]{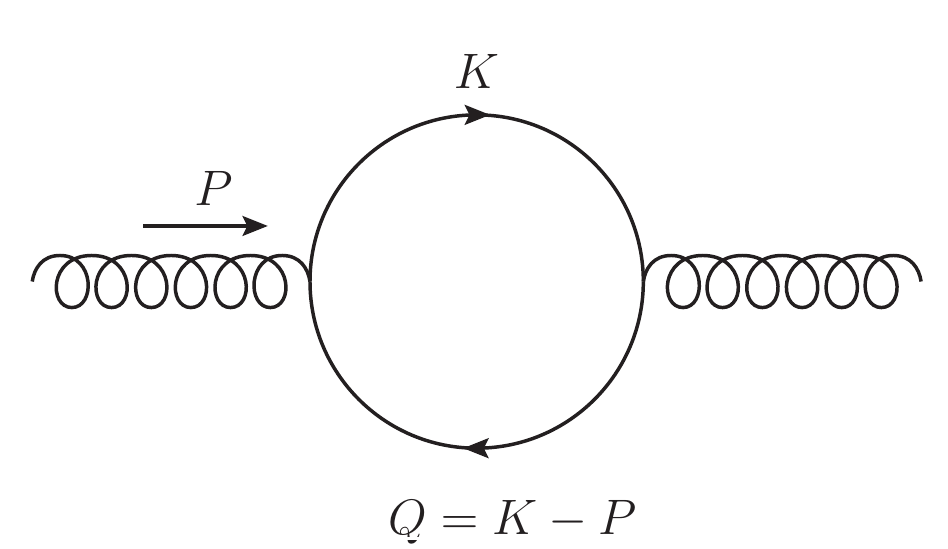}
\caption{The quark loop is the only contribution to the gluon self-energy
that is affected by the finite chemical potential $\mu$.}
\label{eq:quark_loop}
\end{figure}

Consider first the gluon self-energy $\Pi^{ab}_{\mu\nu}$. The color
structure of this tensor is trivial, so that
$\Pi_{\mu\nu}^{ab}=\delta^{ab}\Pi_{\mu\nu}$. We have four diagrams to
consider: the quark loop, the gluon loop, the gluon tadpole, and the gluon
ghost. Only the quark loop
is affected by the presence of a chemical potential and
hence we make it the focus of our attention. With all momenta labeled,
this
diagram is shown in figure \ref{eq:quark_loop}.

Our strategy is to evaluate this diagram in the imaginary-time, invert it
to get the
gluon propagator, and then analytically continue the result to retarded
frequencies to get the real-time $ra$ propagator from which all other
propagators in the $r,a$ basis may be obtained. Proceeding, we first get:
\begin{align}
\Pi_{\mu\nu}(P_E) &=g^{2}\Tr(T^{a}T^{a})\int
\{dK_E\}\
Tr(\gm{\mu}S_{F}^{\prime(\mu)}(K_E)\gm{\nu}S_{F}^{\prime(\mu)}(Q_E))
\end{align}
In the above, the integration measure is
$\{dK_E\}=T\sum_{r=-\infty}^{\infty}\int(d\vk)$, where
$(d\vk)=\frac{d^{3}\vk}{(2\pi)^3}$ is the spatial momentum integration
measure. Since we have factored out the color structure of the gluon self-energy,
there is no sum over $a$ in the $\Tr(T^{a}T^{a})$ factor.
Further, $K_E=(i\omega_r,\vk)$ is the fermionic loop momentum and
$Q_E=K_E-P_E=(i\omega_r-i\omega_n,\vk-\vp)$ is the difference between the
momentum of the gluon ($P_E=(i\omega_n,{\bf p})$)
and the virtual quark. The frequency $i\omega_n=2\pi
n T$ is bosonic, and the frequency $i\omega_r=2\pi(r+\frac{1}{2})T$ is
fermionic. The color matrices $T^{a}$ are the generators of
$\mathrm{SU(3)}$
in the fundamental representation. In this representation, the group
factor
$\Tr(T^{a}T^{a})$ is simply $\frac{1}{2}$. $S_{F}^{\prime(\mu)}$ denotes
the fermion propagator time evolved using the ``thermodynamics
Hamiltonian'' $\hH-\mu\hat{Q}$
since this is what satisfies the periodic boundary
condition (see Appendix A for KMS conditions at finite $\mu$).

The cleanest way to compute the above is to use the spectral
representation
\begin{align}
S_{F}^{\prime(\mu)}(i\omega_r,\vk)
&=\int_{0}^{\beta}d\tau\;
e^{i\omega_r\tau}\int_{-\infty}^{\infty}\frac{dk^0}{2\pi}(1-n_f(k^0-\mu))\slashed{K}e^{-\tau(k^0-\mu)}\rho_0(K)
\end{align}
Indeed, this expansion of the propagator makes the frequency sum trivial.
After performing the spin trace and employing standard identities
involving
the Fermi-Dirac distribution, we obtain:
\begin{align}
\Pi^{(a)}_{\mu\nu}(i\omega_n,\vp)&=
2g^2N_f\int(d\vk)\int\frac{dk^0}{2\pi}\int\frac{dq^0}{2\pi}\rho_0(K)\rho_0(Q)\times\nonumber\\
&\quad\times(K_\mu Q_\nu+Q_\mu K_\nu-g_{\mu\nu}K\cdot
Q)\frac{n_f(k^0-\mu)-n_f(q^0-\mu)}{k^0-q^0-i\omega_n}
\end{align}
For the spatial component of the self-energy tensor, the result is
\begin{align}
\Pi_{ij}(i\omega,\vp)&=
 \nonumber 2g^2N_{f}\int(d\vk)\frac{1}{4E_{\vk}E_{\vq}}\biggl[(k_i q_j +
q_i k_j +\delta_{ij}(E_{\vk} E_{\vq}-\vk\cdot \vq))\nonumber\\
&\qquad\times\left(\frac{n_{f}(E_{\vk}+\mu)-n_{f}(E_{\vq}+\mu)}{i\omega+E_{\vk}-E_{\vq}}-\frac{n_{f}(E_\vk-\mu)-n_f(E_\vq-\mu)}{i\omega-E_\vk+E_\vq}\right)\nonumber\\
& \quad +(k_i q_j+q_i k_j-\delta_{ij}(E_\vk E_\vq+\vk\cdot\vq))\nonumber\\
&\qquad\times\left(\frac{1-n_f(E_\vk+\mu)-n_f(E_\vq-\mu)}{i\omega+E_\vk+E_\vq}-\frac{1-n_f(E_\vq+\mu)-n_f(E_\vk-\mu)}{i\omega-E_\vk-E_\vq}\right)\biggr]\label{eq:spatial_gluon__pol_quark_loop}
\end{align}
using
$\rho_{0}(K) = \frac{\pi}{E_\vk} \left( \delta(k^0-k)-\delta(k^0+k)
\right)$.
The above expression holds for arbitrary external momenta. Henceforth, we
only consider the hard thermal loop approximation so that $p\ll k$ after
the analytic continuation $i\omega\mapsto p^0+i\epsilon$ has been
performed.
It turns out that the self-energies in the hard thermal loop
approximation are also valid for arbitrary external momenta $P$ at the one
loop order \cite{Flechsig:1995ju,Kraemmer:1989dr}. Hence, we will
freely use our results as the self-energies of the gluons and quarks when
estimating the size of their propagators.


Using the fact that $p\ll k$, we obtain:
\begin{align}
\Pi_{ij}(p^0,\vp)
&= -g^2 N_f \int(d\vk)\biggl[v_i v_j
\frac{\vv\cdot\vp}{p^0-\vv\cdot\vp+i\epsilon}\frac{d}{dk}(n_f(k-\mu)+n_f(k+\mu))\nonumber\\
&\quad+(v_i v_j-\delta_{ij})\frac{1}{k}(n_f(k-\mu)+n_f(k+\mu))\biggr]
\end{align}
where ${\bf v} = {\bf k}/|{\bf k}|$.
After an integration by parts and an application of the result
\begin{align}
\int_{0}^{\infty}(dk)k\left(\frac{1}{e^{\beta(k+\mu)}+1}+\frac{1}
{e^{\beta(k-\mu)}+1}\right)=\frac{\pi^2 T^2}{6}+\frac{\mu^2}{2}
\end{align}
we get:
\begin{align}
\Pi_{ij}(p^0,\vp)&=
-g^2N_f\left(\frac{\mu^2}{2}+\frac{\pi^2 T^2}{6}\right)
\int {d\Omega\over 8\pi^3}
\left[v_i v_j-\delta_{ij}-2(v_i v_j)
\frac{\vv\cdot\vp}{p^0-\vv\cdot\vp+i\epsilon}\right]
\end{align}
By symmetry, we have $\int d\Omega v_i v_j = \frac{1}{3}\int d\Omega
\delta_{ij}$. Thus, we finally obtain the hard thermal loop part of the
gluon polarization tensor.
\begin{align}
\Pi_{ij}(P) &= g^2 N_f \left(\frac{\mu^2}{2}+\frac{\pi^2
T^2}{6}\right)\int\frac{d\Omega}{4\pi^3}\frac{p^0v_i
v_j}{p^0-\vv\cdot\vp+i\epsilon}
\end{align}

By going through the same analysis, we can evaluate the other components
of
the gluon polarization tensor. If we include the contributions of the
three
other loops, we finally have:
\begin{align}
\Pi_{\mu\nu}(P) &=
m_D^2\left(-\delta_{0}^{\mu}\delta_{0}^{\nu}+\int\frac{d\Omega}{4\pi}\frac{p^0v_\mu v_\nu}{p^0-\vv\cdot\vp+i\epsilon}\right)\label{eq:HTL_gluons}
\end{align}
where in the above, $m_D^2=g^2\left(\frac{N_f T^2}{6}
+\frac{N_c T^2}{3}+\frac{N_f\mu^2}{2\pi^2}\right)$ is the Debye mass, and
$v_\mu=(1,\vv)$.

\begin{figure}[t]
\centering
\includegraphics[width=0.7\textwidth]{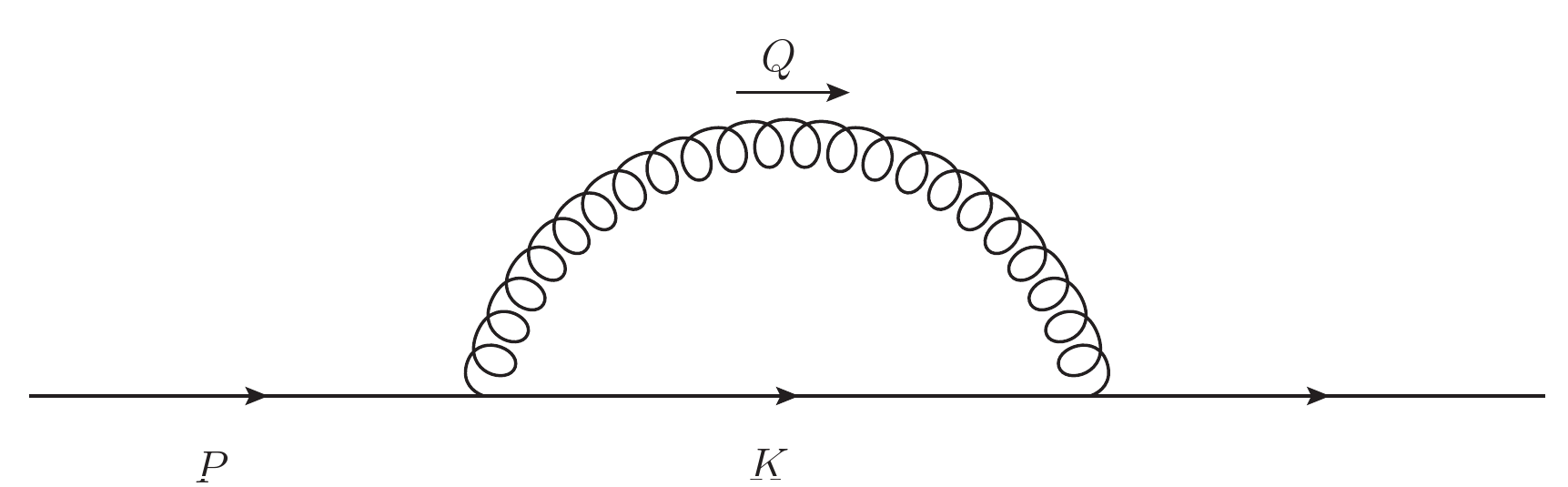}
\caption{The chemical potential appears explicitly in the quark
propagators, whence we need to resum the self-energy of the quark.}
\label{fig:quark_self_energy}
\end{figure}

The contribution of the quark loop to the gluon self-energy corresponds to
the screening of the strong interaction by quarks in the medium. It is
therefore natural to expect that the chemical potential of the quarks will
have an influence on the repeated scattering events that occur during the
photon emission process. The chemical potential also appears explicitly in
the quark propagator and thus affects the thermal mass of the quarks. This
will affect the integral equation that we will derive for photon
production, so we turn to computing the self-energy of the quarks. The
relevant diagram is shown in figure \ref{fig:quark_self_energy}.

The expression corresponding to the diagram in
figure~\ref{fig:quark_self_energy}
is:
\begin{align}
\label{eq:quark_self_energy}\Sigma(P_E)&=
g^2T^aT^a\int\{dQ_E\}\gm{\mu}S_{F}^{\prime\mu}(P_E-Q_E)\gm{\nu}G^{\mu\nu}(Q_E)
\end{align}

In the above, $\int\{dQ_E\}\equiv T\sum_{n=-\infty}^{\infty}\int(d\vq)$
denotes
a sum over bosonic frequencies followed by an integration over the spatial
part of $Q$. As usual, $T^{a}T^{a}=C_2(F)\mathbf{1}=\frac{4}{3}\mathbf{1}$
where $\mathbf{1}$ is the identity matrix in $\mathrm{SU(3)}$. In the
Coulomb gauge, we find:
\begin{align}
\Sigma(i\omega_n, {\bf p}) 
&=-g^2
C_{2}(F)\int\{dQ\}\bigg(\gm{0}S_{F}^{\prime(\mu)}(K_E)\gm{0}
\left(-\frac{1}{q^2}+\int\frac{dq^0}{2\pi}
\frac{\rho_L(q^0,\vq)}{q^0-i\omega_n}\right)\nonumber\\
&\qquad\qquad\qquad
+\gm{i}S_{F}^{\prime(\mu)}(K_E)\gm{j}(\delta_{ij} -\hq_i\hq_j)
\left(\int\frac{dq^0}{2\pi}\frac{\rho_T(q^0,\vq)}{q^0-i\omega_n}\right)\bigg)
\end{align}
where we have used the spectral representation of the longitudinal and
transverse part of the Coulomb gauge gluon propagator following
\cite{Blaizot:2001nr}.  Here $\hq_i = q_i/q$.
The above shows explicitly that we have three
contributions to the quark self energy: $\Sigma(P)=
\Sigma_{C}(P)+\Sigma_L(P)+\Sigma_T(P)$.

The term $\Sigma_C(P)$ is the contribution of the $1/q^2$ term in the
longitudinal piece of the gluon propagator. This corresponds to the
instantaneous Coulomb interaction term. Explicitly, it takes the form:
\begin{align}
\Sigma_C(P_E) &=
g^2C_2(F)\int\{dQ_E\}\frac{1}{q^2}\gm{0}S_{F}^{\prime(\mu)}(K_E)\gm{0}
\end{align}
Once again, using the spectral representation of the fermion propagator,
we
obtain:
\begin{align}
\Sigma_C(P) &=
-g^2C_2(F)\int(d\vq)\frac{1}{q^2}\int_{-\infty}^{\infty}\frac{dk^0}{2\pi}(1-n_f(q^0-\mu))\rho_0(K)\gm{0}\slashed{K}\gm{0}
\end{align}
This has no term proportional to $g^2 T^2$. Thus, we will not pursue
its computation any further, but we will rather focus on the transverse
part. As for the longitudinal part, $\rho_{L}=0$ at tree level, so that it
gives a vanishing contribution -- This is not true for dressed
propagators,
and in particular, $\rho_{L}$ does contribute to the quark damping rate.
After employing the spectral representation of the fermionic propagators,
we obtain for the transverse self-energy:
\begin{align}
\Sigma_T(P_E) 
&= -g^2
C_{2}(F)\int(d\vq)\int\frac{dk^0}{2\pi}
\int\frac{dq^0}{2\pi}(\delta_{ij}-\hq_i\hq_j)
(\gamma^i \slashed{K} \gamma^j)\rho_{0}(K)\rho_{T}(q^0,\vq)\times
\nonumber\\
&\qquad\qquad\qquad\qquad
\times\frac{1+n_b(q^0)-n_f(k^0-\mu)}{k^0+q^0-(i\omega_r+\mu)}
\end{align}

Retarded boundary conditions are obtained from the above by taking the
analytic continuation $i\omega_r+\mu\mapsto p^0+i\epsilon$. Simplifying the
Dirac structure and applying the
hard thermal loop approximation as in the gluon self-energy calculation,
we get:
\begin{align}
\Sigma_{T}(P)
&=\frac{g^2
C_{2}(F)}{8}\left(T^2+\frac{\mu^2}{\pi^2}\right)\int\frac{d\Omega}{4\pi}\frac{\slashed{v}}{p^0-\vv\cdot\vp+i\epsilon}
\end{align}
with $v^\mu=(1,\vv)$.



%

\end{document}